\documentclass[a4paper,11pt]{article}
\usepackage{pos}

\title{New measurements on diffractive vector mesons}

\author[]{Simone Ragoni$^{a,*}$, for the ALICE and LHCb collaborations}

\affiliation[a]{University of Birmingham,\\
  B15 2TT, Birmingham, United Kingdom (UK)}

             \begin{NoHyper}
               \renewcommand\thefootnote{$\speakerFlag$}%
               \footnotetext{Speaker}%
               \stepcounter{footnote}%
               \let\thefootnote\oldthefootnote%
             \end{NoHyper}

\emailAdd{simone.ragoni@cern.ch}

\usepackage{lineno}
\usepackage{xspace}
\usepackage{xcolor}
\usepackage{hyperref}
\usepackage{subfigure}
\usepackage{lineno}
\newcommand{\pp}           {pp\xspace}

\newcommand{\XeXe}         {\mbox{Xe--Xe}\xspace}
\newcommand{\PbPb}         {\mbox{Pb--Pb}\xspace}
\newcommand{\AAtwo}           {\mbox{A--A}\xspace}

\newcommand{\nineH}        {$\sqrt{s}~=~0.9$~Te\kern-.1emV\xspace}
\newcommand{\seven}        {$\sqrt{s}~=~7$~Te\kern-.1emV\xspace}
\newcommand{\eight}        {$\sqrt{s}~=~8$~Te\kern-.1emV\xspace}
\newcommand{\thirteen}     {$\sqrt{s}~=~13$~Te\kern-.1emV\xspace}
\newcommand{\twoH}         {$\sqrt{s}~=~0.2$~Te\kern-.1emV\xspace}
\newcommand{\twosevensix}  {$\sqrt{s}~=~2.76$~Te\kern-.1emV\xspace}
\newcommand{\five}         {$\sqrt{s}~=~5.02$~Te\kern-.1emV\xspace}
\newcommand{\fivefourfour} {$\sqrt{s}~=~5.44$~Te\kern-.1emV\xspace}
\newcommand{\fiveExactly}  {$\sqrt{s}~=~5$~Te\kern-.1emV\xspace}
\newcommand{\twosevensixnn}{$\sqrt{s_{\mathrm{NN}}}~=~2.76$~Te\kern-.1emV\xspace}
\newcommand{\fivenn}       {$\sqrt{s_{\mathrm{NN}}}~=~5.02$~Te\kern-.1emV\xspace}

\newcommand{\GeVc}         {Ge\kern-.1emV/$c$\xspace}
\newcommand{\MeVc}         {Me\kern-.1emV/$c$\xspace}
\newcommand{\TeV}          {Te\kern-.1emV\xspace}
\newcommand{\GeV}          {Ge\kern-.1emV\xspace}
\newcommand{\MeV}          {Me\kern-.1emV\xspace}
\newcommand{\GeVmass}      {Ge\kern-.2emV/$c^2$\xspace}
\newcommand{\MeVmass}      {Me\kern-.2emV/$c^2$\xspace}

\newcommand{\jpsi} {\ensuremath{{\mathrm J}/\psi}\xspace}

\newcommand{\rhozero} {\ensuremath{\rho^0}\xspace}
\newcommand{\twohundrednn}       {$\sqrt{s_{\mathrm{NN}}}~=~200$~Ge\kern-.1emV\xspace}

\abstract{Vector meson photoproduction at high energies has attracted increased interest in recent years due to the unique kinematical ranges offered by the LHC experiments. The ALICE and LHCb collaborations have provided measurements of vector meson photoproduction on nuclear and proton targets, i.e. off p, Pb and Xe targets. The large data sets available from LHC Run 2 allow for more differential measurements. 

The ALICE and LHCb collaborations have recently provided  measurements of coherent \jpsi in \PbPb ultra-peripheral collisions (UPC). In addition, the former also reports the first measurement of the $|t|$-dependence of coherent \jpsi in \PbPb, and has also recently measured coherent \rhozero photoproduction off Pb and Xe targets, thus providing the first study of the A dependence of coherent \rhozero photoproduction.

Vector meson photoproduction in peripheral \PbPb collisions has also been observed by ALICE and LHCb, thus providing a viable technique to extract photonuclear cross sections at a low Bjorken-$x$ of around $10^{-5}$. Lastly, the potential of UPC for tetraquark searches is briefly discussed.
}

\FullConference{%
 
 The Ninth Annual Conference on Large Hadron Collider Physics - LHCP2021\\
 7-12 June 2021\\
 Online
}

\begin{document}
\maketitle

\section{Introduction}
\begin{figure}[ht!]
  \begin{minipage}[c]{0.6\textwidth}
    \includegraphics[width=1.\textwidth]{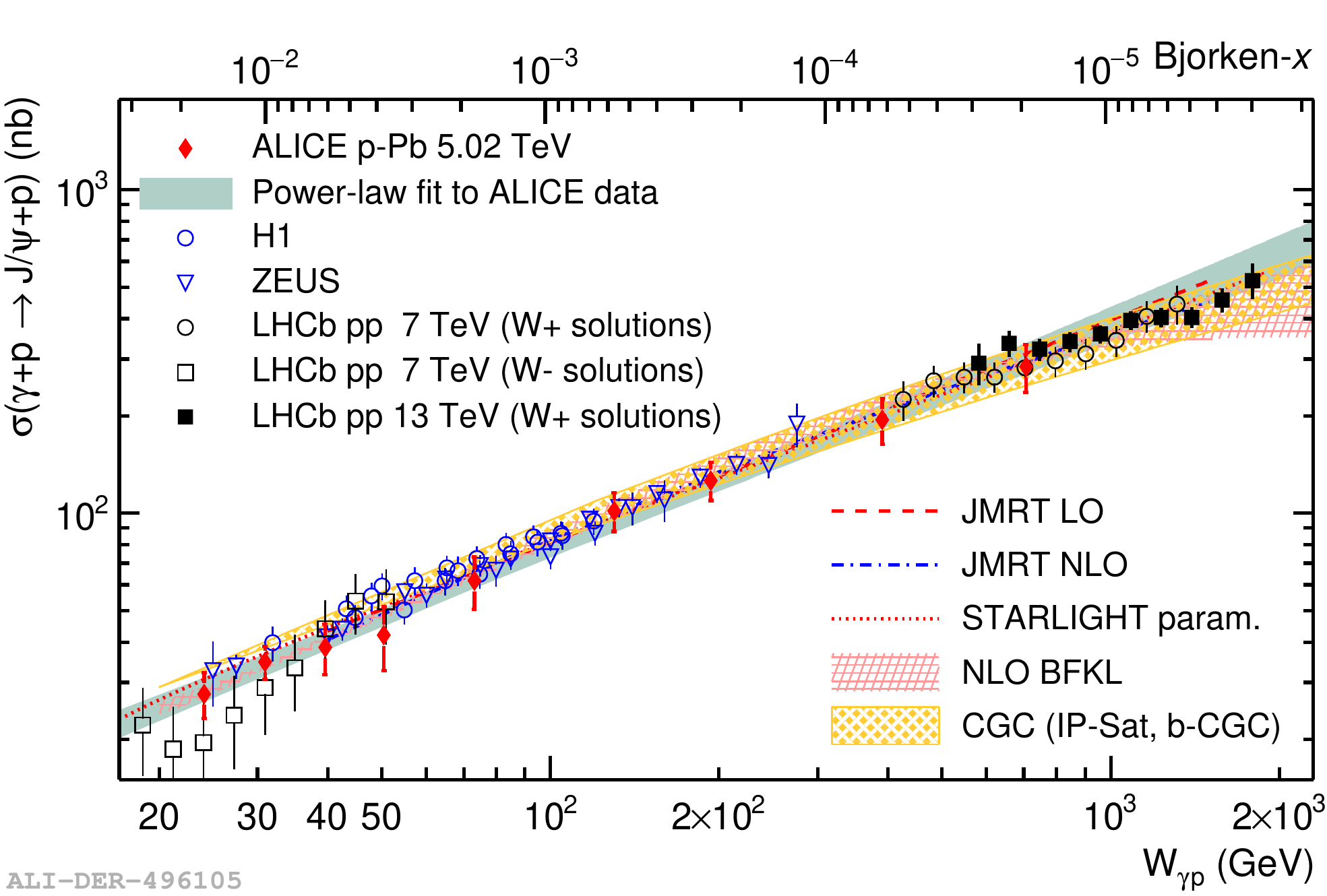}
  \end{minipage}\hfill
  \begin{minipage}[c]{0.38\textwidth}
    \caption{
       Exclusive \jpsi production cross section as a function of the centre-of-mass energy of the $\gamma$p system \cite{Aaij:2018arx, TheALICE:2014dwa,  Acharya:2018jua, ALICE:2018oyo}. 
    } \label{fig:peripheral}
  \end{minipage}
\end{figure}

The LHC experiments offer new and hitherto unexplored kinematical ranges for the investigation of vector meson photoproduction, thus renewing the interest in these processes. The ALICE, ATLAS, CMS, and LHCb collaborations all report measurements for ultra-peripheral collisions, and this contribution will focus on the results on vector meson photoproduction from ALICE and LHCb.  The main motivation behind this interest, lies in the possibility of studying the gluon distributions at low Bjorken-$x$. The Bjorken-$x$ involved in the process can be directly related to the rapidity of the vector meson in the final state, via $x = \frac{M_{\rm VM}}{\sqrt{s_{\rm NN}}}\cdot \exp(\pm |y|)$ \cite{Contreras:2015dqa}. It is then possible to probe the gluon PDFs down  to  $x{\sim}10^{-6}$ in \pp at LHCb \cite{Aaij:2013jxj, Aaij:2014iea, Aaij:2018arx}, with the coverage in $W_{\gamma {\rm p}}$ available to ALICE and LHCb, as shown in Fig.~\ref{fig:peripheral} for exclusive \jpsi photoproduction \cite{Aaij:2018arx}. 
Both sets of data roughly follow a power law, as previously observed with H1 data \cite{Alexa:2013xxa}. However, at higher energies, the LHCb high $W_{\gamma {\rm p}}$ solutions agree better with the NLO JMRT prediction, which deviates somewhat from power law.

\section{Coherent \jpsi photoproduction in \PbPb collisions}
\begin{figure}[ht!]
  \begin{minipage}[c]{0.6\textwidth}
    \includegraphics[width=0.8\textwidth]{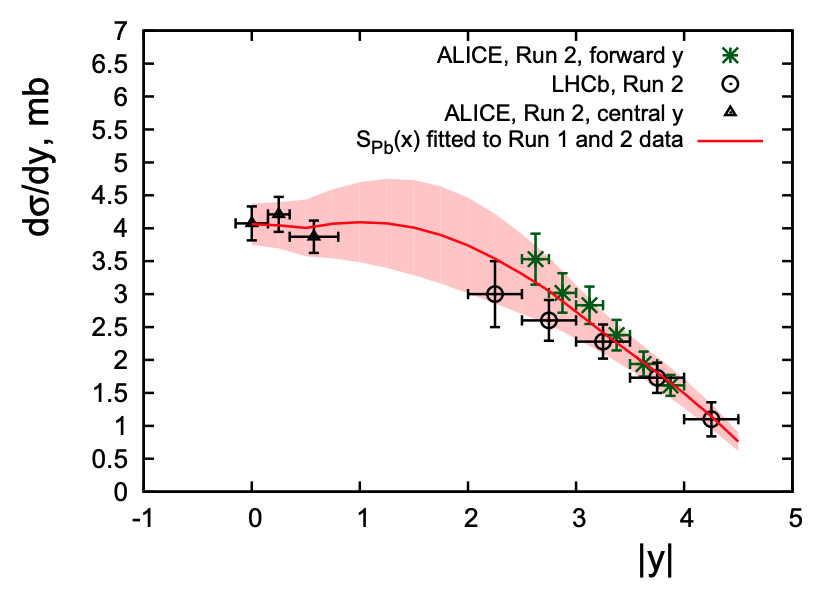}
  \end{minipage}\hfill
  \begin{minipage}[c]{0.38\textwidth}
    \caption{
       Fit to the ALICE and LHCb measurements for coherent \jpsi photoproduction in \PbPb UPCs \cite{Guzey:2020ntc}. The ALICE and LHCb cross sections agree within current systematical uncertainties. The fit favours a $S_{{\rm Pb}(x) } \sim 0.6$ with a 5\% accuracy at $x = 6\cdot 10^{-4}-10^{-3}$.
    } \label{fig:alice-lhcb}
  \end{minipage}
\end{figure}

Coherent \jpsi photoproduction has been measured in \PbPb ultra-peripheral collisions (UPC) by the ALICE and LHCb collaborations \cite{Acharya:2019vlb, ALICE:2021gpt, Aaij:2021jsl}. In UPCs, the two Pb ions are at large impact parameters $b$, greater than the sum of the two nuclear radii, i.e. $b > 2\cdot R_{\rm Pb}$, where $R_{\rm Pb}$ is the radius of the Pb ion. A possible process is when a photon from one nucleus interacts coherently with the other nucleus, resulting in a vector meson in the final state. The nuclei remain largely intact. The final state is thus especially clean: only the decay products of the vector meson are observed, in an otherwise empty detector. The main purpose of coherent \jpsi photoproduction is to deepen  understanding of nuclear shadowing. The nuclear suppression factor, $S_{{\rm Pb}(x) }$, is an interesting variable to gauge the nuclear shadowing involved in a measurement \cite{Guzey:2020ntc}. It is defined starting from the measured cross section, as shown in the following Eq.~\ref{eq:nuclear-suppression-factor} \cite{Guzey:2020ntc}:
\begin{equation}
    \label{eq:nuclear-suppression-factor}
    S_{{\rm Pb}(x)} = \sqrt{\frac{\sigma(\gamma A \longrightarrow \jpsi A)_{measured}}{\sigma(\gamma A \longrightarrow \jpsi A)_{\rm IA}}} \text{ ,}
\end{equation}
where $\sigma(\gamma A \longrightarrow \jpsi A)_{measured}$ is the measured cross section for coherent \jpsi photoproduction, while $\sigma(\gamma A \longrightarrow \jpsi A)_{\rm IA}$ is the cross section computed for the Impulse Approximation model, which neglects nuclear effects. 
The ALICE and LHCb measured cross sections agree within current systematical uncertainties, as shown in Fig.~\ref{fig:alice-lhcb} \cite{Guzey:2020ntc, Acharya:2019vlb, ALICE:2021gpt, Aaij:2021jsl}. The cross sections can be simultaneously fitted to measure the nuclear suppression factor. The data are consistent with $S_{{\rm Pb}(x) } \sim 0.6$ with a 5\% accuracy at $x = 6\cdot 10^{-4}-10^{-3}$\cite{Guzey:2020ntc, ALICE:2021gpt}. The statistical and systematic uncertainties from the measured cross sections are added in quadrature together for the fit.

Ultimately, the small uncertainties on $S_{{\rm Pb}(x) }$ demonstrate the potential of UPC measurements for low-$x$ constraints. Already they resulted in a significant reduction of the uncertainties on EPPS16 nuclear PDFs \cite{Guzey:2020ntc}. 

\section{Extracting photonuclear cross sections in nucleus--nucleus collisions }
Vector meson photoproduction in \AAtwo collisions suffers from an intrinsic ambiguity, due to the process being a superposition of two diagrams, i.e. the system is symmetric. This implies that the cross section ${\mathrm d}\sigma_{\rm PbPb}/{\mathrm d}y$ has two contributions, as shown in  Eq.~\ref{eq:xsec}:
\begin{equation}
 \label{eq:xsec}
    {\mathrm d}\sigma_{\rm PbPb}/{\mathrm d}y = n(\gamma, +|y|)\cdot \sigma_{\gamma {\rm Pb}}(+|y|) + n(\gamma, -|y|)\cdot \sigma_{\gamma {\rm Pb}}(-|y|)\text{ ,}
\end{equation}
where $n(\gamma, \pm|y|)$ represents the photon flux as a function of rapidity, and $\sigma_{\gamma {\rm Pb}}(\pm|y|)$ the photonuclear cross section.
It is possible to disentangle the two photonuclear cross sections by measuring cross sections at different impact parameters. Two techniques have been considered, namely \textit{peripheral} photoproduction \cite{Contreras:2016pkc}, and photoproduction accompanied by \textit{neutron emission} \cite{Guzey:2013jaa}. The strategy for the latter is to tag an event based on the neutron activity seen by Zero Degree Calorimeters. UPC events accompanied by neutron emission are characterised by smaller average impact parameters \cite{Broz:2019kpl}. Peripheral collisions are characterised by an impact parameter $b < 2\cdot R_{\rm Pb}$. The ALICE \cite{Bugnon:2020vti}, STAR \cite{STAR:2019yox}, and LHCb collaborations \cite{LHCb:2021hoq} have observed coherent \jpsi photoproduction  in peripheral collisions.

\section{Coherent \rhozero and A dependence}
\begin{figure}[ht!]
     \begin{center}
        \subfigure[]{
            \label{fig:rho}
            \includegraphics[width=0.4\textwidth]{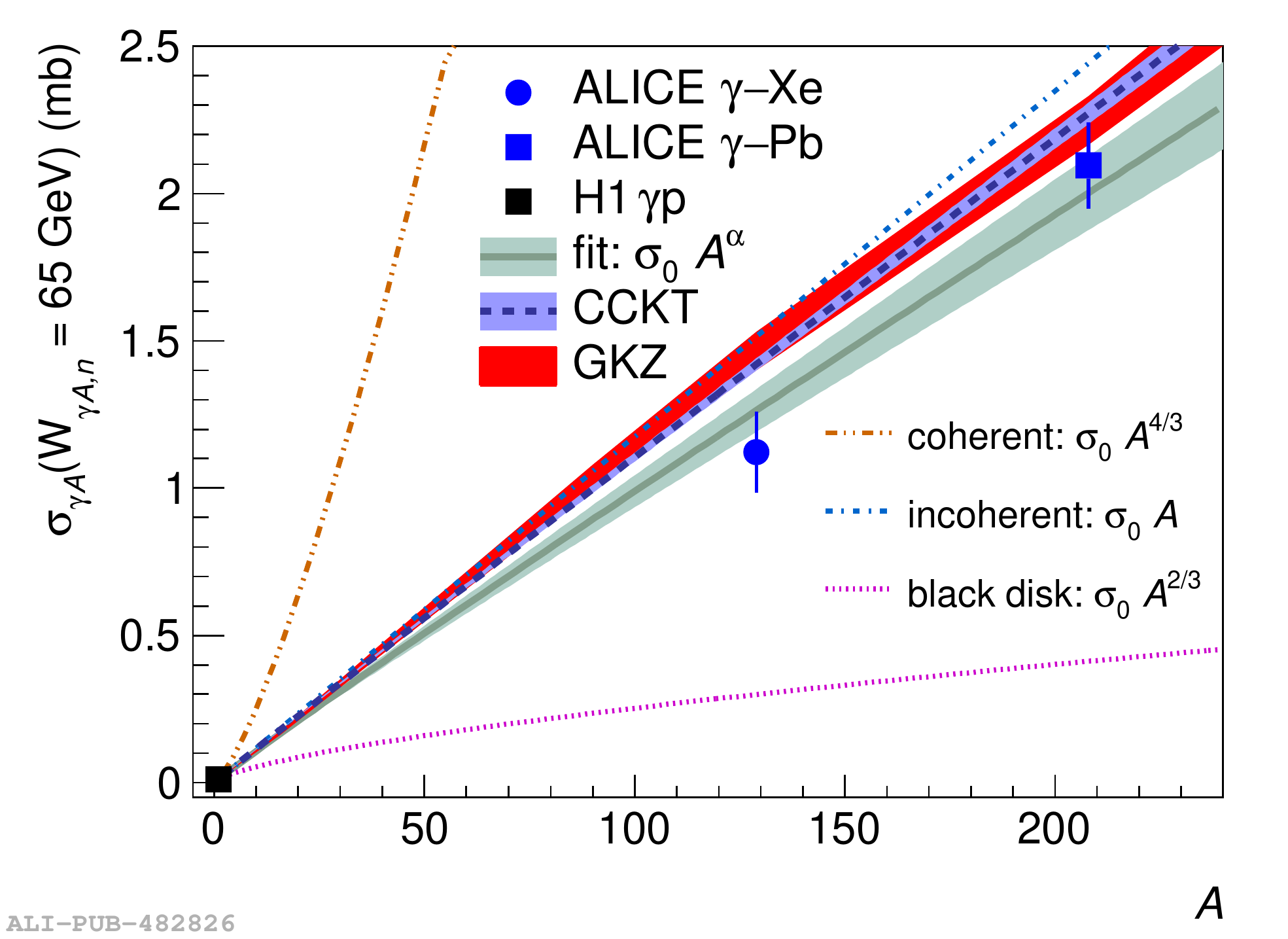}
        }
        \subfigure[]{
            \label{fig:tetra}
            \includegraphics[width=0.4\textwidth]{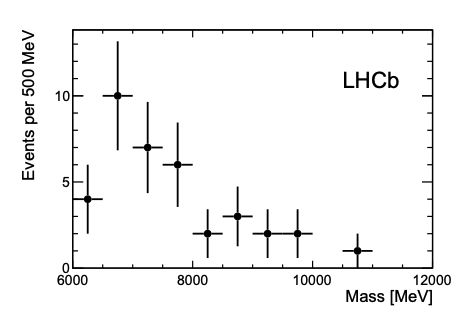}
        }\\
      \end{center}
  \caption{Left: fit to the available coherent \rhozero photoproduction measurements. Nuclear effects are found to be important \cite{ALICE:2021jnv}. Right: effective mass distribution for \jpsi\jpsi pairs in central exclusive production (CEP) events, i.e. \pp$\longrightarrow $\pp$ + X \longrightarrow$\pp$ + \jpsi + \jpsi$  with the LHCb detector  \cite{LHCb:2014zwa}.}
  \label{fig:rho-tetra}
\end{figure}

The ALICE Collaboration has measured coherent \rhozero photoproduction in \PbPb \cite{Acharya:2020sbc} and \XeXe collisions \cite{ALICE:2021jnv}. It is thus now possible to perform the first fit to the available measured cross sections for $\gamma$p, $\gamma$Pb, and $\gamma$Xe, to measure the A dependence of the coherent \rhozero cross section. 
This is shown in Fig.~\ref{fig:rho}, where the fit is performed. A power-law model of the form  $\sigma_{\gamma {\rm A}} = \sigma_0 \cdot {\rm A}^\alpha$, is used to model the trend. The results are compared to three extreme scenarios for the slope $\alpha$: 4/3, 1, and 2/3, for coherent, incoherent and black-disk limit, respectively. The slope is found to be quite close to 1, which does not signify incoherent behaviour but it is  due to large nuclear shadowing suppression  \cite{ALICE:2021jnv}.
\section{Tetraquark searches and UPC}
The LHCb Collaboration has recently observed a structure in the di-\jpsi invariant mass spectrum with inclusive \pp data, $X(6900)$  \cite{LHCb:2020bwg}. This resonance could originate from a  tetraquark state with four charm quarks, i.e. $T_{cc\bar{c}\bar{c}}$, although further investigations are required to ultimately confirm the nature of the resonance. 

It was recently argued \cite{Goncalves:2021ytq} that UPCs might prove to be an interesting opportunity to investigate $T_{QQ\bar{Q}\bar{Q}}$ states, where $Q = c, b$. The processes $\gamma\gamma \longrightarrow \jpsi\jpsi$ and $\gamma\gamma \longrightarrow \Upsilon\Upsilon$ could provide further insights for possible tetraquark states. The LHCb Collaboration had observed the production of \jpsi pairs in exclusive processes, i.e. \pp$ \longrightarrow $\pp$ + X \longrightarrow $\pp$ + \jpsi + \jpsi$ with LHC Run 1 data \cite{LHCb:2014zwa}. As shown in Fig.~\ref{fig:tetra}, the few observed candidates lie in the interesting invariant mass region of the $X(6900)$ resonance. 

\section{Conclusions}
Vector meson photoproduction at LHC energies offers the possibility of addressing gluon distributions at low Bjorken-$x$ down to $x \sim 10^{-6}$. The available measurements on coherent \jpsi photoproduction better constrain the uncertainties on the nuclear suppression factor.  The first fit to the available measurements of coherent \rhozero photoproduction finds significant nuclear effects. Finally, UPCs offer new opportunities for tetraquark investigations with the ALICE and LHCb detectors.

\bibliographystyle{JHEP}
\bibliography{bibliography}

\end{document}